\documentstyle[epsfig]{aipproc}

\begin{document}

\title{The detection of linear polarization in the afterglow of 
GRB~990510 and its theoretical implications}

\author{Davide Lazzati$^{*\dagger}$, Stefano Covino$^*$ and Gabriele 
Ghisellini$^*$}

\address{$^*$ Osservatorio Astronomico di Brera, via E. Bianchi 46, 
I--23807 Merate (LC) Italy \\
$^\dagger$ Dipartimento di Fisica, Universit\`a degli Studi di Mailano,
via Celoria 16, I--20133 Milano, Italy}

\maketitle

\begin{abstract}
We present the recent discovery of linear polarization of the optical 
afterglow of GRB~990510. Effects that could introduce spurious polarization 
are discussed, showing that they do not apply to the case of GRB~990510, 
which is then intrinsically polarized. It will be shown that this 
observation constrains the emission mechanism of the afterglow radiation,
the geometry of the fireball and degree of order of the magnetic field. 
We then present the theoretical interpretations of this observation with 
particular emphasis on the possibility of observing polarization in beamed 
fireballs.
\end{abstract}

\section*{Introduction}

Polarization is one of the clearest signatures of synchrotron radiation, if 
this is produced by electrons gyrating in a magnetic field that is at least 
in part ordered. For this reason, polarization measurements can provide a 
crucial test of the synchrotron shock model \cite{mes97}, the leading
scenario for the production of the burst and, in particular, 
the afterglow photons.

Attempts to measure the degree of linear polarization yelded only
an upper limit ($\sim 2.3\%$ for GRB~990123\cite{hjo99}), until
the observations on the afterglow of the burst of May 10, 1999.
A small but significant amount of polarization was detected
($1.7\pm0.2\%$\cite{cov99}) $\sim 18$ hours after the BATSE trigger
and confirmed in a subsequent observation two hours later\cite{wij99}.

Even if synchrotron radiation can naturally account for the presence of
linearly polarized light in a GRB afterglow, a significant degree of 
anisotropy in the magnetic field configuration or in the fireball
geometry is required. If, in fact, the synchrotron emission is produced
in a fully symmetrical set-up, all the polarization components average 
out giving a net unpolarized flux. The presence of partially
ordered magnetic field (in causally disconnected domains) has been
discussed by Gruzinov \& Waxman \cite{gru99}, however their model
overpredicts, in its simplest formulation, 
the observed amount of polarization. Here we discuss a
different possibility, in which the asymmetry is provided by a collimated
fireball observed off--axis, while the magnetic field is tangled
in the plane perpendicular to the velocity vector of the fireball expansion.
Indeed, the smooth break in the lightcurve of GRB~990510 \cite{isr99} 
has been interpreted as due to a collimated fireball observed slightly
off--axis.

\section*{GRB~990510 measurements}

GRB~990510 was detected by BATSE on-board the Compton Gamma Ray Observatory
and by the {\it Beppo}SAX Gamma Ray Burst Monitor and Wide Field Camera  
on 1999 May 10.36743 UT \cite{kip99,dad99}. Its fluence (2.5$\times 
10^{-5}$ erg cm$^{-2}$ above 20 keV) was relatively high \cite{kip99}. 
Follow up optical observations started $\sim 3.5$~hr later and revealed 
an $R\simeq 17.5$ \cite{axe99} optical transient (OT).
The OT showed initially a fairly slow flux decay $F_\nu\propto t^{-0.85}$
\cite{isr99}, which gradually steepened; Vreeswijk et al. \cite{vre99}
detected Fe II and Mg II absorption lines in the optical spectrum of the 
afterglow. This provides a lower limit of $z=1.619\pm 0.002$ to the redshift, 
and a $\gamma$--ray energy of $> 10^{53}$~erg, in the case of isotropic 
emission.

\begin{figure}[tb!]
\parbox{7.5truecm}{\epsfig{file=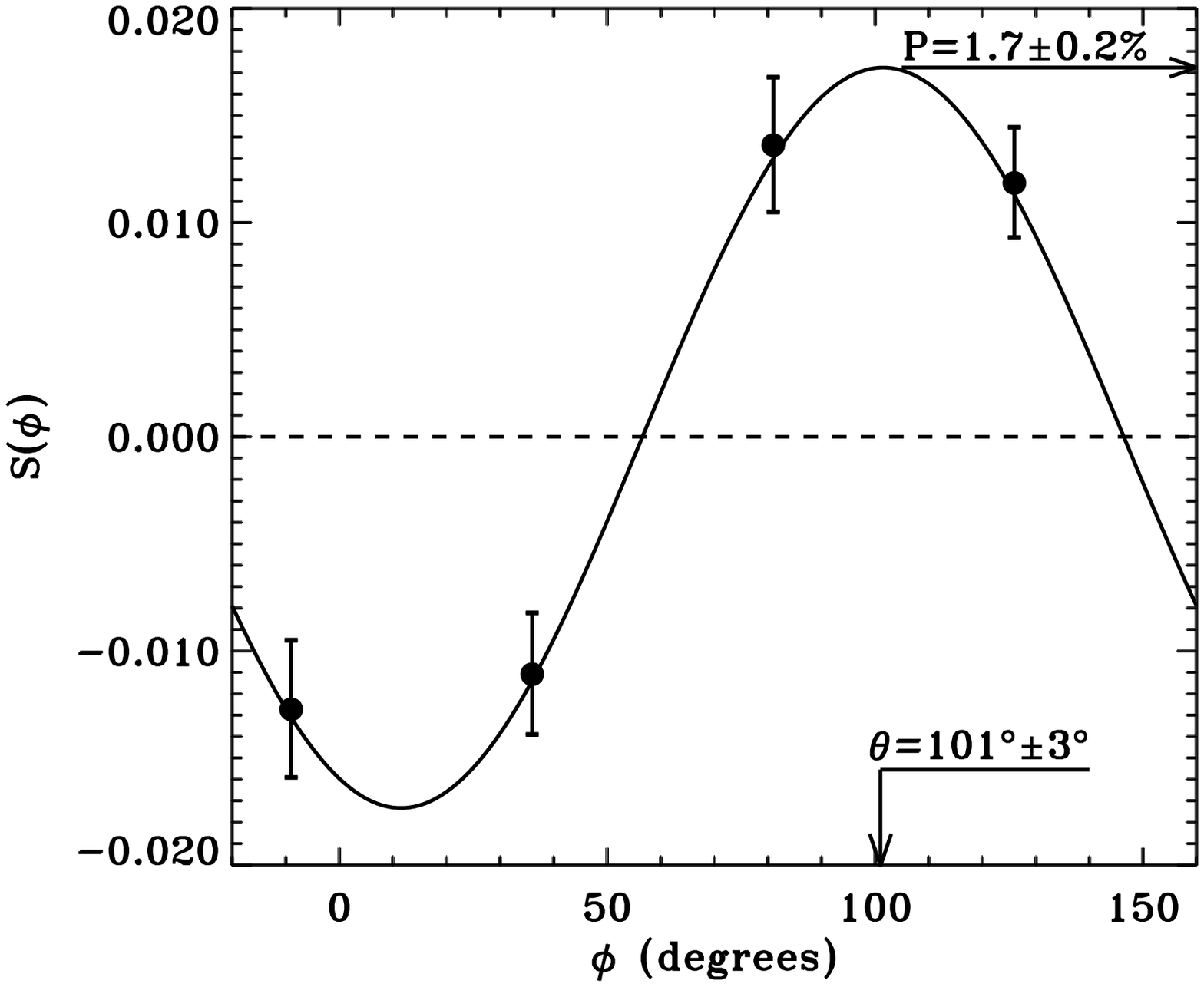,width=7.2truecm}}
\parbox{6.5truecm}{\epsfig{file=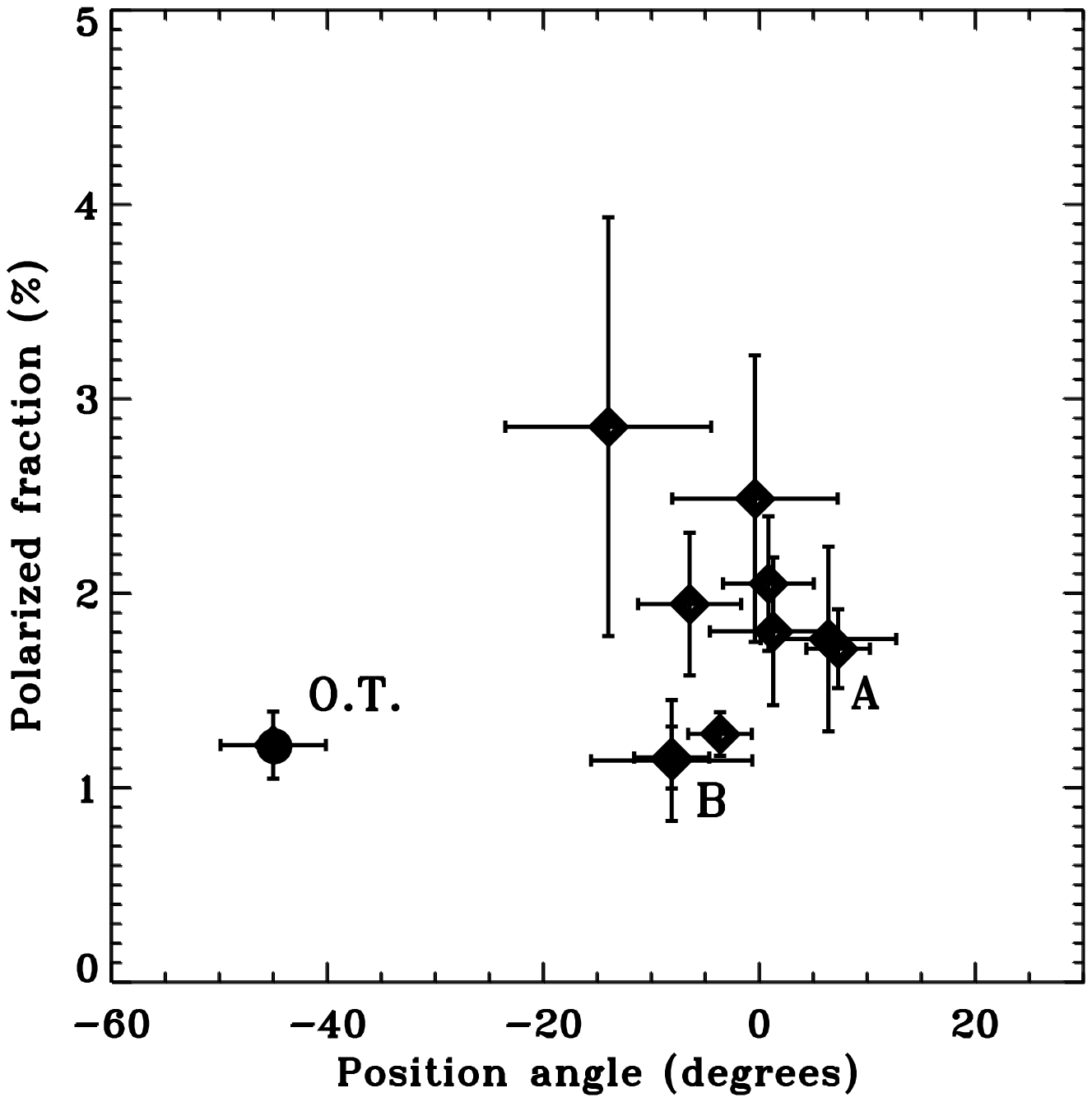,width=6.2truecm}}
\vspace{10pt}
\caption{{{\bf Left Panel}: our polarization data taken at four different 
position angles $\phi$ are fitted with a cosine curve. The amplitude of this 
curve corresponds to the degree of linear polarization, and its maximum 
to the polarization position angle. Data are normalized to the average 
of the stars in the same field. {\bf Right Panel}:
The unnormalized 
degree of polarization vs. the instrumental polarization 
position angle of the stars in the field and the optical transient.
The optical transient clearly stands out (P = $1.2 \pm 0.2$\%). 
The two stars closest to the OT are labelled A and B.
}
\label{fig1}}
\end{figure}

We observed the OT associated with GRB~990510 $\sim 18$ hours after the 
gamma--ray trigger at the ESO VLT-Antu (UT1) in polarimetric mode, 
performing four 10 minutes exposures in the R band at four angles 
($0^\circ$, $22.5^\circ$, $45^\circ$ and $67.5^\circ$) of the 
retarder plate\cite{cov99}. The average magnitude of the OT in the 
four exposures was $R\sim 19.1$.
Relative photometry with respect to all the stars in the field 
was performed and each couple of simultaneous measurements at orthogonal 
angles was used to compute the points in Fig.~\ref{fig1} (laft panel)
(see\cite{cov99} for details). 
The parameter $S(\phi)$ is related to the degree of linear
polarization $P$ and to the position angle of the electric field vector 
$\vartheta$ by:
\begin{equation}
S(\phi)\, =\, P \cos 2(\vartheta - \phi).
\end{equation}
$P$ and $\vartheta$ are evaluated by fitting a cosine curve to the observed 
values of $S(\phi)$. The derived linear polarization of the OT of GRB 990510 
is $P=(1.7\pm 0.2)$\% (1$\sigma$ error), at a position angle of  
$\vartheta=101^\circ \pm 3^\circ$.
Fig.~\ref{fig1} (left panel) shows the data points and the 
best fit $\cos\phi$ curve. 
The statistical significance of this measurement is very high. 
A potential problem is represented by a ``spurious'' polarization introduced 
by dust grains interposed along the line of sight, which may be preferentially 
aligned in one direction.
The normalization of the OT measurements to the stars 
in the field already corrects for the average interstellar 
polarization of these stars, 
even if this does not necessarily account 
for all the effects of the galactic ISM along the line of sight to the OT
(e.g. the ISM could be more distant than the stars, not
inducing any polarization of their light).
To check this possibility, we plot in Fig.~\ref{fig1} (right panel) the 
degree of polarization vs. the instrumental position 
angle for each star and for the OT.
It is apparent that, while the position angle of all stars are consistent 
with being the same (within 10 degrees), the OT clearly stands out.
The polarization position angle of stars close to the OT
differs by $\sim 45^\circ$ from the position angle of the OT.
This is contrary to what one would expect if the polarization of the OT 
were due to the galactic ISM. 
Polarization induced by absorption in the host galaxy can be constrained to be 
$P_{host} < 0.2\%$, due to the lack of any absorption in the optical filters
in addition to the local value (see\cite{cov99} for more details).
We therefore conclude that the OT, even if contaminated by interstellar 
polarization, must be intrinsically polarized to give the observed orientation.

\section*{Polarization from beamed fireballs}

\begin{figure}[thb!]
\centerline{\epsfig{file=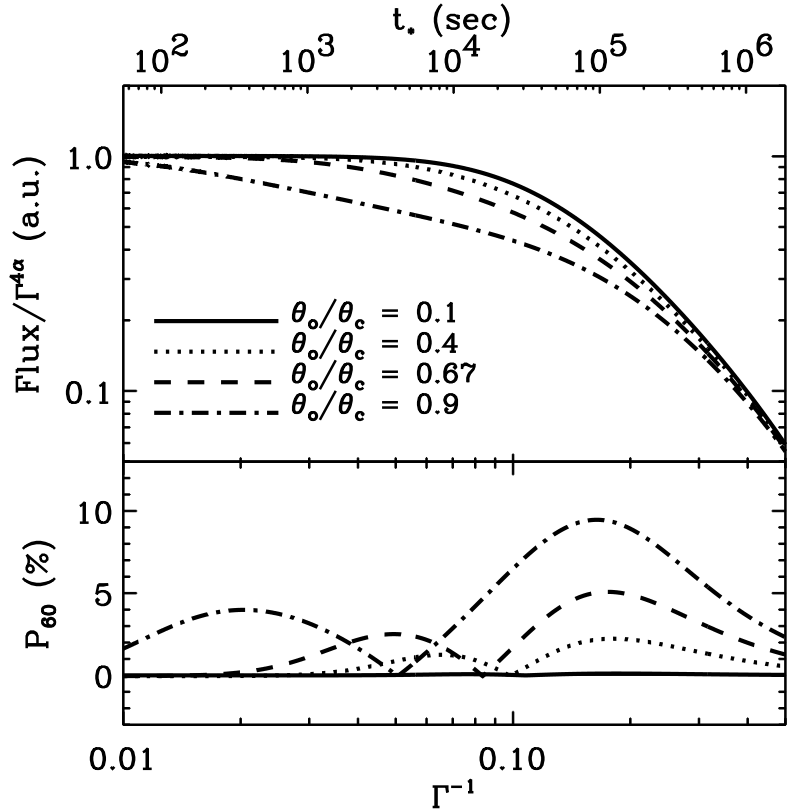,width=10truecm}}
\vspace{10pt}
\caption{{Lightcurves of the total flux (upper panel) and of the
polarized fraction (bottom panel) for four different choices
of the ratio $\theta_o/\theta_c$. The cone aperture angle $\theta_c=5^\circ$, 
while $\theta_o$ is the viewing angle.
The higher the ratio $\theta_o/\theta_c$, the higher the polarized fraction
due to the increase of the asymmetry of the geometrical setup.
The actual value of the observed polarization depends linearly
upon $P_0$. In this figure we assumed $P_0=60$\%.
The lightcurve of the total flux assumes a constant spectral index
$\alpha=0.6$ for the emitted radiation. 
Note that the the highest polarization values are associated with
total flux lightcurves steepening more gently.
To calculate the values of the upper x--axis (t$_\star$), we assumed
$(t_\star/t_0) = (\Gamma/\Gamma_0)^{-8/3}$ with
$t_0=50$ s and $\Gamma_0=100$.}
\label{fig2}}
\end{figure}

We consider a slab of magnetized plasma, in which the configuration of the
magnetic field is completely tangled if the slab is observed face on,
while it has some some degree of alignment if the slab is observed edge on.
Such a field can be produced by compression in one direction of a volume
of 3D tangled magnetic field\cite{lai80} or by Weibel 
instability\cite{med99}.
If the slab is observed edge--on, the radiation is therefore polarized at a 
level, $P_0$, which depends on the degree of order of the field in the plane. 
If the emitting slab moves in the direction normal to its plane
with a bulk Lorentz factor $\Gamma$, we have to take into
account the relativistic aberration of photons.
This effect causes photons emitted at $\theta^\prime=\pi/2$ in the (primed) 
comoving frame $K^\prime$ to be observed at $\theta\sim 1/\Gamma$ 
(see also\cite{med99}).

We assume that the fireball is collimated into a cone 
of semi--aperture angle $\theta_c$, and that the line of sight 
makes an angle $\theta_o$ with the jet axis.
As long as $\Gamma>1/(\theta_c-\theta_o)$, the observer receives photons 
from a circle of semi-aperture angle $1/\Gamma$ around $\theta_o$.
Consider the edge of this circle: radiation coming from each sector is highly
polarized, with the electric field oscillating in radial direction 
(see\cite{ghi99} for more details).
As long as we observe the entire circle, the configuration is
symmetrical, making the total polarization to vanish.
However, if the observer does not see part of the circle, some net 
polarization survives in the observed radiation. 
This happens if a beamed fireball is observed off--axis when
$1/(\theta_c+\theta_o)<\Gamma<1/(\theta_c-\theta_o)$.

At the beginning of the afterglow, when $\Gamma$ is large, the observer 
sees only a small fraction of the fireball and no polarization is observed.
At later times, when $\Gamma$ becomes smaller than $1/(\theta_c-\theta_o)$,
the observer will see only part of the circle centered in $\theta_o$:
there is then an asymmetry, and a corresponding net polarized flux.
To understand why the polarization angle in this configuration is horizontal, 
consider that the part of the circle which is not observed would have 
contributed to the polarization in the vertical direction. 
At later times, as the fireball slows down even more, a larger area becomes 
visible. When $\Gamma\sim 1/(\theta_c+\theta_o)$, the dominant contribution
to the flux comes from the upper regions of the fireball
which are vertically polarized.
The change of the position angle happens when the contributions from 
horizontal and vertical polarization are equal, resulting in a vanishing 
net polarization.
At still later times, when $\Gamma \to 1$, light aberration
vanishes, the observed magnetic field is completely tangled
and the polarization disappears.

Figure~\ref{fig2} shows the result of the numerical integration of the
appropriate equations (see\cite{ghi99} for the detailed discussion).
As derived in the above qualitative discussion, the lightcurve of the
polarized fraction shows two maxima, with the position angle
rotated by 90$^\circ$ between them. It is interesting to note the link
with the lightcurve. The upper panel of Fig.~\ref{fig2} shows the lightcurve
of the total flux divided by the same lightcurve in the assumption
of spherical geometry. As expected, the lightcurve of the beamed fireball
shows a break with respect to the spherical one.
A larger off--axis ratio produce a more gentle
break in the lightcurve, and is associated with a larger value
of the polarized fraction. The behaviour of the total flux and of the 
polarization lightcurves allow us to constrain the off--axis
ratio $\vartheta_o/\vartheta_c$, but is insensitive to the absolute
value of the beaming angle $\vartheta_c$. Therefore, even if we could 
densely sample the polarization lightcurve, the beaming angle could be
derived only assuming a density for the interstellar medium,
i.e. a relation between the observed time and the braking law of the 
fireball.
On the other hand, the detection of a $90^\circ$ rotation of the 
polarization angle of the afterglow would be the clearest sign of 
beaming of the fireball, expecially if associated with a smooth
break in the lightcurve.
Polarimetric follow up of afterglows is hence a powerful
tool to investigate the geometry of fireballs.

\end{document}